\documentclass[preprint,preprintnumbers,amsmath,amssymb]{revtex4}
\usepackage{graphicx}
\usepackage{epsfig}
\usepackage{lscape}
\usepackage{dcolumn}
\usepackage{bm}
\begin{document}
\title{ Comparison of van der Waals coefficient C$_{6}$ of sodium clusters obtained via
spherical jellium background model and all-electron \textit{ab initio method} }
\author{Arup Banerjee$^{a}$ and  Manoj K. Harbola $^{b}$\\
(a) Laser Physics Application Division, Raja Ramanna Centre for Advanced Technology\\
Indore 452013, India\\
(b) Department of Physics, Indian Institute of Technology, Kanpur\\
U.P 208016, India}
\begin{abstract}
In this paper we employ two distinct approaches - all-electron \textit{ab initio} method and the spherical jellium background model- within time dependent density functional theory  to calculate the long range dipole-dipole dispersion coefficient (van der Waals coefficient) $C_{6}$ of sodium atom clusters containing even number of atoms ranging from 2 to 20 atoms. The dispersion coefficients are obtained via Casimir-Polder relation. All the calculations are carried out with local density approximation for exchange-correlation energy functional. These two sets of results are  compared to assess the accuracy of jellium based results and to ascertain the effect of detail ionic structure of the clusters on the van der Waals coefficient.  
\end{abstract}
 \maketitle 
 \section{Introduction}
Contribution of long-range dispersive van der Waals forces to the interaction between two molecules is quite significant. These forces play an important role in the description of many physical and chemical phenomena. Some of these are adhesion, surface tension, physical adsorption \cite{langbein}, chemistry of rare gases \cite{christe}, molecular chemistry in the interstellar medium \cite{flower} etc.. The correlation between the electron density fluctuation at widely separated locations give rise to these forces. The  van Waals interaction between two molecules A and B in the lowest order is represented by the orientationally averaged potential 
\begin{equation}
V_{AB}(R) = -\frac{C_{6}}{R^{6}}
\label{vanderwaalspot}  
\end{equation}
where R is the intermolecular distance and C$_{6}$ is the orientationally averaged van der Waals coefficient. The coefficient C$_{6}$ describes the dipole-dipole interactions between the two molecular species.

Metal clusters, specifically those of alkali atoms Na and K, played an important role in the development of cluster physics as a branch of modern physics and chemistry. The knowledge of their van der Waals coefficient $C_{6}$ is useful for the description of cluster-cluster collisions \cite{amadon} and also for characterizing the orientation of clusters in the bulk matter \cite{gunnarsson,lambin}. However, only very few papers devoted to the calculation of van der Waals coefficients  exist in the literature. In Refs. \cite{pacheco1,pacheco2}, time dependent Kohn-Sham (TDKS) equation of time dependent density functional theory (TDDFT) within the spherical jellium background model (SJBM) was employed to calculate the van der Waals coefficients. On the other hand, in Ref. \cite{banerjee}, a purely particle- and current-density based hydrodynamic formalism of TDDFT within SJBM has been applied to calculate the coefficients. 
This model is well suited for the description of the alkali metal clusters (Li, Na, and K) and correctly explains the greater stability of these clusters with magic number 2, 8, 20, 34, 40, $\cdots$ of atoms due to closing of the electronic shell \cite{brackreview,alonso}.
Although the hydrodynamic approach is less accurate than the TDKS approach due to the approximate nature of the kinetic energy functional employed in it, nonetheless it is computationally advantageous for larger clusters in comparison to the orbital-based TDKS approach.  

The SJBM replaces the discrete ionic structure of clusters by a spherically symmetric uniform positive charge background and thus making it possible to carry out calculations for the optical response properties of reasonably large clusters of around 100 atoms within the TDKS approach \cite{madjet}. In contrast to this the hydrodynamic approach in conjunction with SJBM allows calculation of optical response properties of very large alkali atom clusters containing 10,000 atoms without much difficulty \cite{brack,banerjee1,banerjee,harbola}. In past ten years or so, several all-electron \textit{ab initio}  calculations devoted to the ground state and the optical response properties of sodium clusters taking into account the actual geometrical arrangement of the sodium atoms have also been reported in the literature  \cite{martins,moulett,andreoni,guan,calmanici,kummel,kronik,blundell,pacheco3,solovyov,ghanty}. However, these calculations are restricted to small sized clusters containing maximum up to 20 sodium atoms. Moreover, these calculations are computationally more expensive and require substantially more resources than the jellium based calculations. It is only very recently that the all-electron \textit{ab initio} calculations of the van der Waals coefficient C$_{6}$ of small sized closed shell sodium clusters containing up to 20 atoms have been reported in the literature \cite{jiemchooroj,banerjee2}. The main aim of this paper is to make a systematic comparison of the values of C$_{6}$ obtained by employing the jellium based model and all-electron \textit{ab initio} method within the TDDFT. This comparison will enable us to assess the accuracy of jellium model in predicting the values of van der Waals coefficient and such a study is important, as the jellium model often turns out to be much more efficient to handle particularly when dealing with larger cluster systems. 

Before proceeding further, it is important to note that
density functional theory (DFT) in principle should give the exact
ground-state properties including the long range van der Waals energies.
However, the widely used local density approximation (LDA) and generalized gradient approximations (GGA)
\cite{gga1,gga2,gga3} exchange-correlation (XC) functionals fail to reproduce the van der Waals
energies. This is because the LDA and GGA XC functionals cannot
completely simulate the correlated motion of electrons arising
from Coulomb interaction between distant non overlapping
electronic systems. It is only recently that attempts
\cite{andersson,dobson,kohn} have been made to obtain van der
Waals energies directly from the ground-state energy functional by
correcting the long range nature of the effective Kohn-Sham
potential. On the other hand, it is possible to make reliable
estimates of the van der Waals coefficient $C_{6}$ by using expressions
which relate this coefficient to the frequency dependent
dipole polarizabilities at imaginary frequencies
\cite{casimirpolder,stone}. We follow the latter route for the calculation of 
these coefficients.

The paper is organized as follows: In section II we discuss the theoretical method and the expressions employed to calculate the van der Waals coefficient $C_{6}$ from the frequency dependent dipole polarizability. Results of our calculations are presented in Section III. The paper is concluded in Section IV.

\section{Method of Calculation} 
In order to calculate the van der Waals coefficient $C_{6}$, we make use of the Casimir-Polder expression which relates $C_{6}$ to the frequency dependent dipole polarizability evaluated at imaginary frequency. In accordance with this expression the orientation averaged dispersion coefficient between two molecules A  and B is given by \cite{casimirpolder,stone}
\begin{equation}
C_{6}(A,B) = \frac{3}{\pi}\int_{0}^{\infty}d\omega\bar{\alpha}_{A}(i\omega
)\bar{\alpha}_{B}(i\omega ) \label{casimirpolder}
\end{equation}
where $\bar{\alpha}_{j}(i\omega)$ is the isotropic average dipole polarizability of the j-th molecule and is given by
\begin{equation}
\bar{\alpha}_{j}(\omega) = \frac{\alpha^{j}_{xx}(\omega) + \alpha^{j}_{yy}(\omega) + \alpha^{j}_{zz}(\omega)}{3}.
\end{equation}
In the above expression $\alpha_{xx}(\omega)$, $\alpha_{yy}(\omega)$ and $\alpha_{zz}(\omega)$ are diagonal elements of the dipole polarizability tensor. Therefore, the calculation of dispersion coefficient $C_{6}$ involves determining frequency dependent dipole polarizability tensor followed by the evaluation of the quadrature. For the determination of the frequency dependent polarizability, we use two different methods (i) the hydrodynamic formalism of TDDFT in conjunction with the SJBM and (ii) all-electron \textit{ab initio} method also in the realm of TDDFT. For details, we refer the reader to Refs. \cite{harbola,banerjee1,banerjee}  for the hydrodynamic formalism and Ref. \cite{gisbergen} for the \textit{ab initio} method of calculation.  In the following, we describe the choice of parameters made in this paper to perform the calculations of the dispersion coefficient C$_{6}$ by two methods. 
\subsection{Hydrodynamic-SJBM Formalism}
The basic idea of the SJBM is to replace the distribution of ionic cores by a constant spherically symmetric positive background of jellium density $n_{+}({\bf r})$ in a finite volume. This positive background provides the attractive potential for the valence electrons contributed by each atom of the cluster. The ground-state electronic distribution is then obtained by employing either orbital based Kohn-Sham or purely density based extended Thomas-Fermi (ETF) formalisms of DFT.  The calculation of the frequency dependent polarizabilty within the SJBM can be accomplished by TDDFT. Like its time independent counterpart TDDFT can be formulated either in terms of the time dependent orbitals (TDKS formalism) or in terms of  the particle and the current densities (hydrodynamic formalism). In this paper, we employ the hydrodynamic approach formalism of TDDFT in conjunction with the SJBM to calculate the
frequency dependent polarizabilities at imaginary frequencies. The basic dynamical variables of the hydrodynamic theory are the time-dependent density $\rho ({\bf r},t)$ and the velocity
potential $S({\bf r},t)$. The velocity of the electron fluid is given by ${\bf v}({\bf r},t)= -{\bf\nabla}S({\bf r},t)$.  Thus the total time-averaged energy can be
expressed in terms of these two variables. For our purpose we need to evaluate
the second-order change in the time-averaged energy as this is
related directly to the frequency dependent average dipole
polarizability by the relation
\begin{equation}
\bar{\alpha}(\omega ) = -4E^{(2)}(\omega) \label{eq12}
\end{equation}
where $\omega$ is the frequency of the applied electromagnetic field. It has been shown that the second-order change in the time-average energy $E^{(2)}(\omega)$ is stationary with respect to the variations in the first-order induced density $\rho^{(1)}({\bf r},\omega)$ and the
induced current-density $S^{(1)}({\bf r},\omega)$ \cite{banerjee1}. Consequently,
$E^{(2)}(\omega)$ can be determined by choosing appropriate
variational forms for $\rho^{(1)}({\bf r},\omega)$ and $S^{(1)}({\bf
r},\omega)$ and making $E^{(2)}$ stationary with respect to the
parameters of $\rho^{(1)}({\bf r},\omega)$ and $S^{(1)}({\bf r},\omega)$. For our calculations we use 
ten parameters each for $\rho^{(1)}({\bf r},\omega)$ and $S^{(1)}({\bf r},\omega)$ and check the convergence by adding more parameters. The variational method
is also applicable for the imaginary frequencies \cite{banerjee} as it requires replacing
$\omega^{2}$ by $-\omega^{2}$ in the expression of $E^{(2)}(\omega)$. This allows us to
determine dynamic multipolarizability at imaginary frequencies
($\alpha (i\omega)$) by exactly the same procedure as employed for
getting $\alpha (\omega)$.

It is well known that a TDDFT based response property calculation requires approximating the XC functional at two different levels. The first one is the static XC potential needed to calculate the ground-state density of the system. The second approximation is needed to represent the XC kernel $f_{XC}({\bf r},{\bf r'},\omega)$ which determines the XC contribution to the screening of an applied field. For the XC kernel, we use reasonably accurate adiabatic local density approximation (ALDA) \cite{gross}. On the other hand, for static XC energy functional we use  the Dirac exchange energy functional \cite{dirac} and  Gunnarsson-Lundqvist (GL) \cite{gl} parametrized form for the correlation energy functional within the LDA.
In addition to approximating the XC energy functional, the calculation of polarizability by the hydrodynamical approach of TDDFT also requires
approximating the non-interacting kinetic energy functional $T_{s} [\rho ]$ also. We choose the von Weizsacker \cite{weizsacker} form for
$T_{s} [\rho ]$, which is given as
\begin{equation}
T_{W}[\rho] =
\frac{1}{8}\int\frac{{\bf\nabla}\rho\cdot{\bf\nabla}\rho}
{\rho}d{\bf r}. \label{14}
\end{equation}
The von Weiszacker functional is well suited for the description of response properties of closed shell metal clusters \cite{banerjee}.  

In the present paper the ground-state densities $\rho^{(0)}({\bf
r})$ of clusters are obtained by employing purely density-based 
ETF \cite{brackreview,brack} method within the SJBM of metal clusters. 
\subsection{\textit{Ab initio} TDDFT based method}
To carry out the all-electron \textit{ab initio} calculation of the frequency dependent polarizabilities of the clusters, we employ the ADF program package \cite{adf}. We refer the reader to Ref. \cite{gisbergen} for detailed description of the method adopted in this package for obtaining the frequency dependent polarizabilities. In this package calculations of electronic and response properties of molecules are carried by using Slater type orbital (STO) basis sets. It is well known that for accurate calculations of response properties it is necessary to have large basis sets with both polarization and diffuse functions. For our purpose, we choose all electron even tempered basis set ET-QZ3P-2DIFFUSE with two sets of diffuse functions consisting of (11s,9p,7d,3f) functions for Na atom.  The application of basis set with diffuse functions often leads to the problem of linear dependencies. Such problem have been circumvented by removing linear combinations of functions corresponding to small eigenvalues of the overlap matrix. 
We expect that the size of the chosen basis set will make our results very close to the basis-set limit. Once the average polarizability at imaginary frequencies are obtained then the Casimir-Polder integral Eq. (\ref{casimirpolder}) is evaluated by employing thirty point Gauss-Chebyshev quadrature scheme as described in Ref. \cite{rijks}. The convergence of the results has been checked by increasing number of frequency points. Like SJBM based calculations, all the \textit{ab initio} calculations in the present paper are performed with the ALDA for the XC kernel. On the other hand, for the static XC potential needed to calculate the ground-state orbitals and their energies, we employ the LDA potential as parametrized by Vosko, Wilk and Nussair (VWN) \cite{vwn}. We note here that both GL and VWN forms for the XC potential use the same Dirac exchange energy functional \cite{dirac} but the parameterization for the correlation part is different. We expect that this difference in the correlation energy functional will lead to a significantly smaller deviation in the results obtained via the \textit{ab initio} and SJBM calculations than the discrepancy arising due to the consideration of the actual geometrical structure of the clusters in the \textit{ab initio} calculations.
Furthermore, in order to perform \textit{ab initio} calculation of response properties we need to choose the ground-state geometries of clusters. For the dimer $Na_{2}$, we use the experimental bond length 3.0786 $\AA$. On the other hand, for larger clusters (4- to 20-atom clusters) we use the structures which are obtained via geometry optimization calculation with triple-$\xi$ plus two polarization functions (TZ2P) basis set and Becke-Perdew (BP86) XC potential \cite{becke2,perdew2}. All the optimizations are carried out with the convergence criteria for the norm of energy gradient and energy fixed at $10^{-4}$a.u and $10^{-6}$a.u, respectively. The optimized structures obtained by us are in agreement with the corresponding results of Refs. \cite{solovyov,ghanty}. In case of a cluster having more than one isomers we choose the one possessing the lowest energy for our calculations of the dipole polarizability. Here we note that our geometry optimization calculation for the cluster containing 20 sodium atoms yields lower energy for the structure with C$_{2v}$ symmetry than the one with T$_{d}$ symmetry. This is in contrast to the result of Ref. \cite{solovyov}.  The next section is devoted to the discussion of results for the dispersion coefficient $C_{6}$ obtained by employing the above two methods.
\section{Results and Discussion}
\begin{table}[h]
\caption{Comparison of dispersion coefficient $C_{6}$ (in atomic units) for pairs of similar sodium clusters obtained with the all-electron \textit{ab initio} and SJBM calculations.}
\begin{center}
\begin{tabular}{|c|c|c|c|}\hline
$N$ & \multicolumn{3}{c|}{$C_{6}\times 10^{-3}$ a.u. }  \\
\cline{2-4}
& \textit{ab initio} & SJBM-Hydrodynamic & SJBM-KS$^{a}$  \\
\hline
2 & 3.68 & 2.63 & 2.62  \\
4 & 15.05 & 9.73 & -  \\
6 & 29.44 & 20.65 & -  \\
8 & 41.82 & 35.43 & 40.06  \\
10 & 71.11 & 54.04 & -  \\
12 & 103.74 & 76.43 & -  \\
14 & 137.45 & 102.56 & -  \\
16 & 160.38 & 132.38 & -  \\
18 & 191.23 & 165.87 & -  \\
20 & 232.81 & 202.99 & 228.58  \\
\hline
\end{tabular}
\end{center}
(a) Ref. \cite{pacheco2}
\end{table}
\begin{figure}[ht]
\begin{center}
\psfig{file=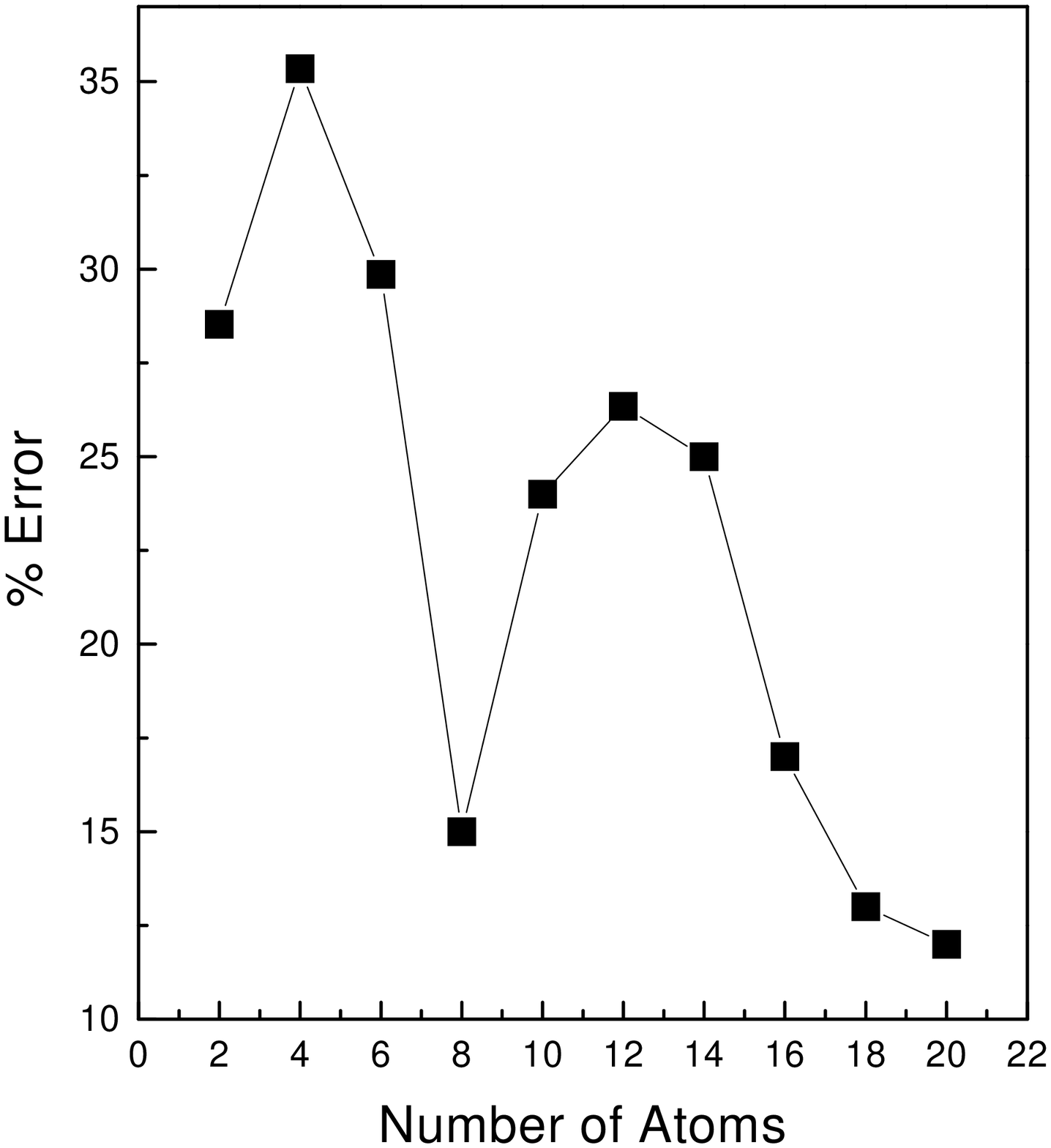,width=3.0in}
\caption{Percent difference $100\times(C_{6}^{\textit{ab initio}} - C_{6}^{SJBM})/C_{6}^{\textit{ab initio}}$ between the \textit{ab initio} and SJBM-hydrodynamic results for the van der Waals coefficients C$_{6}$ as a function of number of atoms. The lines joining the points are guide to the eye. }
\label{fig1}
\end{center}
\end{figure}
We begin this section with a comparison of the results for $C_{6}$ between similar pairs of  sodium clusters (Na$_{n}$-Na$_{n}$) with even number of atoms $n$ ranging from 2 to 20. These results are shown in Table I. For comparison, we also include in Table I the results for C$_{6}$ of 2-, 8-, and 20-atom clusters obtained by employing TDKS formalism within the SJBM \cite{pacheco2}. We note that the values of C$_{6}$ for the dimer obtained within SJBM by employing hydrodynamic and TDKS approaches are almost the same. This is because the von Weizsacker form for the kinetic energy functional is exact for two-electron systems. For 8- and 20-atom clusters the difference between the TDKS and hydrodynamic results are of the order of 10$\%$. In our earlier study \cite{banerjee}, we have found similar order of difference between the two results for 40-atom cluster also. This then demonstrates that the SJBM-hydrodynamic results are reasonably accurate and capable of yielding C$_{6}$ quite close to the more accurate orbital based TDKS number. 
Having assessed the accuracy of the SJBM-hydrodynamic results for the van der Waals coefficient C$_{6}$ with respect to the corresponding TDKS results, we next focus our attention on their comparison with the corresponding \textit{ab initio} numbers. 
First it is evident from Table I that the results based on SJBM are lower than the \textit{ab initio} numbers for all the clusters considered in this paper. The deviation is more for the non-magic number clusters. In order to estimate the deviation between the two results, we plot in Fig. 1 the relative error given by $(C_{6}^{\textit{ab initio}} - C_{6}^{SJBM})/C_{6}^{\textit{ab initio}}$ as a function of the number of atoms. Fig. 1 clearly reveals some important results of this paper. We notice that the relative error shows an oscillatory behaviour with the increasing number of atoms. It is important to note here that the relative errors for the closed shell magic number clusters assume the positions of minima. Moreover, the relative error for these magic number clusters decrease as the number of atoms present in the cluster increases. For example, the relative error for the dimer is about 28$\%$ and it goes down to approximately 12$\%$ for the 20-atom cluster. This is an encouraging results as for the clusters larger than Na$_{20}$ performing \textit{ab initio} calculations of response properties like frequency dependent polarizability is computationally quite demanding. For other clusters the relative error in magnitude with maxima at 4- and 12-atom clusters. However, the value of relative error for the 12-atom cluster is lower than for the 4-atom one. We expect that for larger non-magic number clusters the SJBM results for C$_{6}$ will be closer to the corresponding \textit{ab initio} numbers. 
Encouraged by these results, we perform calculation of C$_{6}$ for all pairs of clusters ($Na_{n}-Na_{n}$). The results of these calculations are presented in Table II. Again it is  seen from Table II that for different pairs of magic-number clusters the relative difference between the \textit{ab initio} and the SJBM results is less as compared to the other pairs. From Table II, we also infer that the jellium based results for clusters larger than the 14-atom one are quite close to the corresponding \textit{ab initio} numbers. These results then demonstrate that for clusters larger than Na$_{14}$ the detailed ionic core structure does not have much influence on the values of the van der Waals coefficient C$_{6}$.   

\begin{table}[h]
\caption{Comparison of dispersion coefficient $C_{6}$ $\times 10^{-3}$ (in atomic units) for all pairs of  sodium clusters obtained with the all-electron \textit{ab initio} and SJBM calculations. The SJBM results are shown in the parenthesis} 
\tabcolsep=0.1in
\begin{center}
\begin{tabular}{ccccccccccc}
N  & 2 & 4 & 6 & 8 & 10 & 12 & 14 & 16 & 18 & 20 \\
\hline
2 & 3.68 & 7.45 & 10.41 & 12.39 & 16.18 & 19.54 & 22.49 & 24.27 & 24.48& 29.22  \\
  & (2.63)& (5.06) & (7.38) & (9.66)& (11.93) & (14.19) & (16.43)& (18.67 )& (20.89)& (23.11) \\  
4 &  & 15.05 & 21,04& 25.02 & 32.67& 39.47 & 45.42 & 48.99& 53.44&63.85 \\
  & & (9.73)& (14.17) & (18.56) & (22.92) & (27.25) & (31.57)& (35.87 )& (40.15)& (44.41)\\ 
6  &    &    & 29.44& 35.05& 45.74 & 55.25 & 63.59& 68.59 & 74.91& 89.40  \\
   &    &    & (20.65) & (27.05) & (33.41) & (39.73) & (46.02)& (52.28)& (58.52)& (64.74)\\
8  &    &    & & 41.82& 54.51 & 65.84 & 75.80& 81.89 & 89.42& 98.66  \\
    &   &     &   &(35.43)&(43.76)& (52.04)&(60.28)&(68.48)&(76.66)&(84.75)\\ 
10  &    &    & & & 71.11 & 85.89 & 98.86& 106.75 & 116.53& 128.57  \\
&    &    & & & (54.04) & (66.26) & (74.44)& (84.57) & (94.67)& (104.73)  \\
12  &    &    & & &  & 103.74 & 119.41& 128.93 & 140.74& 155.28  \\
  &    &    & & &  & (76.43) & (88.53)& (100.59) & (112.59)& (124.55)  \\
14  &    &    & & &  &  & 137.45& 148.43 & 162.03& 178.77 \\
  &    &    & & &  &  & (102.56)& (116.52) & (130.43)& (144.28) \\
16  &    &    & & & &  & & 160.38 & 175.12& 193.22 \\
  &    &    & & & &  & & (132.38)) & (148.18) & (163.93) \\
18 &    &    & & & &  & &  & 191.23 & 211.06 \\
 &    &    & & & &  & &  & (165.87) & (183.50) \\
20 &    &    & & &  &  & &  & & 232.81  \\
 &    &    & & &  &  & &  & & (202.99)  \\
\hline 
\end{tabular}
\end{center}
\end{table}
\newpage
Having assessed the accuracy of results obtained within the SJBM, we now focus our attention on the study of relationship between C$_{6}$ and $\bar{\alpha}(0)$. In the so-called London approximation the van der Waals coefficient C$_{6}$ between two similar molecules can be represented in terms of the static polarizability $\bar{\alpha}(0)$ as
\begin{equation}
C_{6} = \frac{3\omega_{1}}{4}(\bar{\alpha}(0))^{2},
\label{londondispersion}
\end{equation}  
where $\omega_{1}$ is the effective or characteristic frequency of the system. The above expression (Eq. (\ref{londondispersion})) is obtained with the single pole approximation for the frequency dependent polarizability, which assumes that all the transitions can be replaced by only one effective transition. This is also known as London dispersion formula and it provides a way to correlate the van der Waals coefficient with the static polarizability. In doing so the accuracy of C$_{6}$ crucially depends on the precision with which the static polarizability is calculated.
\begin{figure}[b]
\begin{center}
\psfig{file=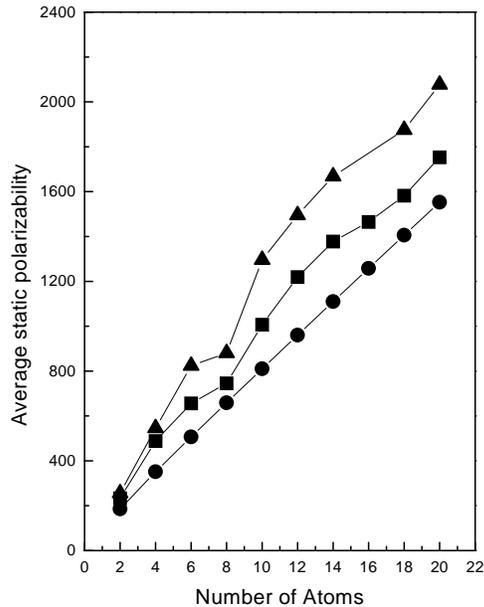,width=3.0in}
\caption{Plot of average static polarizability $\bar{\alpha}(0)$ of sodium atom clusters in atomic units. The solid squares represent the \textit{ab initio} results, SJBM-hydrodynamic results are denoted by solid circles, and solid triangles represent the experimental results of Ref. \cite{knight}. The lines joining the points are guide to the eye. }
\label{fig2}
\end{center}
\end{figure}
In order to study this correlation, we calculate the average static polarizability of sodium clusters by the \textit{ab initio} method and SJBM-hydrodynamic approach. Another reason for carrying out static polarizability calculations is that unlike C$_{6}$, a fairly large number of experimental results on the static polarizabilty exist in the literature \cite{knight,exppol}, which allows us to compare our results with the corresponding experimental data.  
The results of the polarizability calculations are shown in Fig. 2 along with the experimental results of Ref. \cite{knight}. Fig. 2 clearly shows that both \textit{ab initio} and SJBM results are lower than the corresponding experimental data, however, the \textit{ab initio} results are closer to the experimental data than the SJBM results. Moreover, it is seen from this figure that like C$_{6}$, the deviation between the SJBM and \textit{ab initio} results is minimum for magic number clusters and these deviations decrease with the larger clusters. For example, the difference between the two results is 20$\%$ ( \textit{ab initio} number is 233.32 a.u.) for the dimer, whereas the differnce reduces to around 11$\%$ ( \textit{ab initio} number is 1752.8 a.u.) for the Na$_{20}$ cluster. For non-magic number clusters the differnce is more and but reduces with the increase in the size of the clusters. 
Therefore, we conclude that the static polarizability of large size clusters can also be obtained quite accurately by employing the SJBM-hydrodynamic approach with substantially less amount of computational effort. 

\section{Conclusion}
The van der Waals coefficient $C_{6}$ for sodium atom clusters containing even number of atoms ranging from 2 to 20 atoms have been calculated by employing all-electron \textit{ab initio} method and the SJBM-hydrodynamic formalism within the realm of TDDFT. The calculations are performed with the LDA XC potentials. The van der Waals coefficient is obtained by using Casimir-Polder expression which requires frequency dependent dipole polarizabilties of the two interacting species at imaginary frequencies.
These are calculated by the all-electron \textit{ab initio} and the SJBM-hydrodynamic methods of TDDFT. The results of these two calculations are compared to ascertain the effect of ionic structure of the clusters on the van der Waals coefficient. We find that the differences between the results of all-electron \textit{ab initio} and SJBM-hydrodynamic calculations are minimum for magic number clusters and these differences decrease with the increase in the number of atoms present in the clusters. Even for non-magic number clusters, we find that SJBM-hydrodynamic calculations lead to results which are very close to the corresponding \textit{ab initio} numbers. From these results, we conclude that for clusters larger than Na$_{14}$ detailed ionic structure of the clusters does not have much effect on the results for the van der Waals coefficient and also on their static polarizability. This is an important results as carrying out \textit{ab initio} calculations for the optical response properties of large clusters is very expensive. This paper shows that for such systems SJBM based methods within TDDFT can yield quite accurate results for the van der Waals coefficient.

\acknowledgments{We wish to thank Prof. G. Maroulis for inviting us to present our work in this special issue. We also thank Mr. Pranabesh Thander of RRCAT Computer Centre for his help and support in providing us the uninterrupted computational resources and also for smooth running of the codes. It is a pleasure to thank Drs. Aparna Chakrabarti and Tapan K. Ghanty for their help in the geometry optimization and also for useful discussions.}


\clearpage
\newpage 
\begin{thebibliography}{25}
\bibitem{langbein}D. Langbein, {\it Theory of Van der Waals Attraction}, Springer Tracts in Modern Physics Vol. 72 (Springer, Berlin, 1974).
\bibitem{christe}K. O. Christe, Angew. Chem. Int. Ed. {\bf 40}, 1419 (2004).
\bibitem{flower} {\it Molecular Collisions in the interstellar medium}, Astrophysics series {\bf 17}, Ed. D. R. Flower, Cambridge University Press, Cambridge (1990).
\bibitem{amadon}A. S. Amadon and W. H. Marlow, Phys. Rev. A {\bf 43}, 5483 (1991).
\bibitem{gunnarsson} O. Gunnarsson, S. Satpathy, O. Jepsen and O. K. Anderson, Phys. Rev. Lett. {\bf 67}, 3002 (1991).
\bibitem{lambin}Ph. Lambin, A. A. Lucas, J.-P. Vigneron, Phys. Rev. B {\bf 46}, 1794 (1992).
\bibitem{pacheco2}J. M. Pacheco and W. Ekardt, Modern Phys. Lett. B {\bf 7}, 573 (1993)
\bibitem{pacheco1}J. M. Pacheco and W. Ekardt, Phys. Rev. Lett.
{\bf 68}, 3694 (1992).
\bibitem{banerjee} A. Banerjee and M. K. Harbola, J. Chem. Phys.
{\bf 117}, 7845 (2002); A. Banerjee and M. K. Harbola, Proc. Indian Natn. Sci. Acad. {\bf 71 A}, 357 (2005); A. Banerjee and M. K. Harbola, Pramana J. Phys. {\bf 66}, 423 (2006).
\bibitem{brackreview}M. Brack, Rev. Mod. Phys. {\bf 65}, 677 (1993) and references their in; W. A. de Heer, Rev. Mod. Phys. {\bf 65}, 611 (1993).
\bibitem{alonso}J. Alonso and L. C. Balbas {\it Topics in Current Chemistry} {\bf 182}, Ed. R. F. Nalewajski Springer-Verlag Berlin (1996) and references their in.
\bibitem{madjet}M. Madjet, C. Guet, and W. R. Johnson, Phys. Rev. A {\bf 51}, 1327 (1995).
\bibitem{brack}M. Brack. Phys. Rev. B {\bf 39}, 3533 (1989). 
\bibitem{harbola} M. K. Harbola and A. Banerjee, J. Theoretical Comp. Chem. {\bf 2}, 301 (2003).
\bibitem{banerjee1} A. Banerjee and M. K. Harbola, J. Chem. Phys. {\bf 113}, 5614 (2000).
\bibitem{martins}J. L. Martins, J. Buttet, and C. Car, Phys. Rev. B {\bf 31}, 1804 (1985).
\bibitem{moulett} I. Moulett, J. L. Martins, F. Reuse, and J. Buttet, Phys. Rev. B {\bf 42}, 11598 (1990).
\bibitem{andreoni}U. R$\ddot{o}$thlisberger and W. Andreoni, J. Chem. Phys. {\bf 94}, 8129 (1991).
\bibitem{guan}J. Guan, M. E. Casida, A. M. K$\ddot{o}$ster, and D. R. Salahub, Phys. Rev. B. {\bf 52}, 2184 (1995).
\bibitem{calmanici} P. Calmanici, K. Jug, and A. M. K$\ddot{o}$ster, J. Chem. Phys. {\bf 111}, 4613 (1999).
\bibitem{kummel} S. K$\ddot{u}$mmel, T. Berkus, P.-G.Reinhard, M. Brack, Eur Phys. J. D {\bf 11}, 239 (1999).
\bibitem{kronik}L. Kronik, I. Vasiliev, and J. R. Chelikowsky, Phys. Rev. B {\bf 62}, 9992 (2000).
\bibitem{blundell}S. A. Blundell, C. Guet, and R. R. Zope, Phys. Rev. Lett. {\bf 84}, 4826 (2000). 
\bibitem{pacheco3} S. J. A van Gisbergen, J. M. Pacheco, and E. J. Baerends, Phys. Rev. A {\bf 63}, 063201 (2001).
\bibitem{solovyov}I. A. Solovyov, A. V. Solovyov, W. Greiner, Phys. Rev. A {\bf 65}, 053205 (2002).
\bibitem{ghanty}K. R. S. Chandrakumar, T. K. Ghanty, and S. K. Ghosh, J. Chem. Phys. {\bf 120}, 6487 (2004).
\bibitem{jiemchooroj}A. Jiemchooroj, P. Norman and B. E. Sernelius, J. Chem. Phys {\bf 125}, 124306 (2006).
\bibitem{banerjee2} A. Banerjee, A. Chakrabarti, and T. K. Ghanty, accepted for publication in J. Chem. Phys.
\bibitem{gga1}J. P. Perdew, Phys. Rev. Lett. {\bf 55}, 1665
(1985); {\bf 55}, 2370(E) (1985).
\bibitem{gga2} A. D. Becke, Phys. Rev A {\bf 38}, 3098 (1988).
\bibitem{gga3}J. P. Perdew and Y. Wang, Phys. Rev. B, {\bf 33}, 8800 (1986); J. P. Perdew, in {\it Electronic Structure of Solids}, Edited by P. Ziesche and H. Eschrig (Akademic-Verlag, Berlin, 1991).
\bibitem{andersson} Y. Andersson, D. C. Langreth, B. I. Lundqvist, Phys. Rev. Lett. {\bf 76}, 102
(1996).
\bibitem{dobson} J. F. Dobson, B. P. Dinte, Phys. Rev. Lett., {\bf
76}, 1780 (1996).
\bibitem{kohn}W. Kohn, D. Makarov, Phys. Rev. Lett., {\bf 80},
4153 (1998).
\bibitem{casimirpolder}H. B. Casimir and D. Polder, Phys. Rev. {\bf 73}, 360 (1948).`
\bibitem{stone}A. J. Stone, {\it The Theory of Intermolecular Forces}, (Clarendon, Oxford, 1996).
\bibitem{gisbergen}S. J. A. van Gisbergen, J. G. Snijders and E. J. Baerends, J. Chem. Phys. {\bf 103}, 9347 (1995).
\bibitem{gross}E. K. U. Gross, J. F. Dobson and M. Petersilka, in
{\it Density Functional Theory, Topics in Current Chemistry} {\bf
181}, Ed. R. F. Nalewajski (Springer, Berlin, 1996).
\bibitem{dirac}P.A.M. Dirac, Proc. Cambridge Philos. Soc. {\bf 26}, 376 (1930).
\bibitem{gl} O. Gunnarson and B.I. Lundquist, Phys. Rev. B {\bf 13},
4274 (1976).
\bibitem{weizsacker}C. F. von Weizsacker, Z. Phys. {\bf 96}, 431 (1935).
\bibitem{adf} ADF2006.01, SCM, Theoretical Chemistry,
Vrije Universiteit, Amsterdam, The Netherlands, http://www.scm.com.
\bibitem{rijks} W. Rijks and P. E. S. Wormer, J. Chem. Phys. {\bf 88}, 5704 (1988).
\bibitem{vwn}S. H. Vosko, L. Wilk, and M. Nussair, Can. J. Phys. {\bf 58}, 1200 (1980).
\bibitem{becke2} A. D. Becke, Phys. Rev. A {\bf 38}, 3098 (1988).
\bibitem{perdew2} J. P. Perdew, Phys. Rev. B {\bf 33}, 8822 (1986).
\bibitem{knight}W.D. Knight, K. Clemenger, W. A. de Heer, and W. A. Saunders, Phys. Rev. B {\bf 31}, 2539 (1985).
\bibitem{exppol} D. Rayane, A. R. Allouche, E. Benichou,
R. Antoine, M. Aubert-Frecon, Ph. Dugourd, M. Broyer, C. Ristori, F.
Chandezon, B. A. Huber, and C. Guet, Eur. Phys. J. D {\bf 9}, 243 (1999); G. Tikhonov, V. Kasperovich, K. Wong, and V. Kresin, Phys. Rev. A {\bf 64}, 063202 (2001).
\end{thebibliography}
\end{document}